\documentclass[a4paper,12pt]{article}
\pdfoutput=1 

\usepackage{jheppub} 

\usepackage[T1]{fontenc} 

\usepackage{float, array, xspace, amscd, amsthm}
\usepackage{fancyhdr}
\usepackage{longtable}
\usepackage{comment}

\newcommand{\dd}{{\rm d}}
\newcommand{\w}{\wedge}

\newcommand{\IA}{\mathbb{A}}

\newcommand{\IF}{\mathbb{F}}

\newcommand{\cO}{\mathcal{O}}
\newcommand{\cN}{\mathcal{N}}

\newcommand{\be}{\begin{equation}}
\newcommand{\ee}{\end{equation}}
\def\bea#1\eea{\begin{align}#1\end{align}}
\newcommand{\tr}{{\rm tr}}

\newcommand{\Id}{\mathrm{Id}}

\newcommand{\DD}{D} 
\newcommand{\Dtor}{\nabla^{G_2}}
\newcommand{\hM}{\tilde h}

\title{Superpotential of Three Dimensional $\cN=1$ Heterotic Supergravity}
\author[a]{Xenia de la Ossa,}
\author[b]{Magdalena Larfors,}
\author[b]{Matthew~Magill,}
\author[c,d]{Eirik E.~Svanes}

\affiliation[a]{Mathematical Institute, Oxford University\\Andrew Wiles Building, Woodstock Road\\Oxford OX2 6GG, UK }
\affiliation[b]{Department of Physics and Astronomy,Uppsala University\\ SE-751 20 Uppsala, Sweden}
\affiliation[c]{Department of Physics, Kings College London, London, WC2R 2LS, UK}
\affiliation[d]{Abdus Salam International Centre for Theoretical Physics, Strada Costiera 11, 34151, Trieste, Italy}

\emailAdd{delaossa@maths.ox.ac.uk, magdalena.larfors@physics.uu.se, matthew.magill@physics.uu.se, eirik.svanes@kcl.ac.uk}

\abstract{We dimensionally reduce the ten dimensional heterotic supergravity action on spacetimes of the form ${\cal M}_{(2,1)}\times Y$, where ${\cal M}_{(2,1)}$ is three dimensional maximally symmetric Anti de Sitter or Minkowski space, and $Y$ is a compact seven dimensional manifold with $G_2$ structure. In doing so, we derive the real superpotential functional of the corresponding three dimensional ${\cal N}=1$ theory. We confirm that extrema of this functional correspond to supersymmetric heterotic compactifications on manifolds of $G_2$ structure in the large volume, weak coupling limit to first order in $\alpha'$. We make some comments on the role of the superpotential functional with respect to the coupled moduli problem of instanton bundles over $G_2$ manifolds.}

\begin{document}

\maketitle
\flushbottom

\newpage

\section{Introduction}

Heterotic string compactifications have long been understood to offer advantages to the phenomenologist. This is largely due to the fact that a gauge theory is an intrinsic part of the low-energy theory, in contrast to Type II compactifications where manifold singularities and brane intersection patterns are needed to engineer gauge multiplets.  On the downside, the supersymmetry constraints and Bianchi identities (including corrections at order $\cO(\alpha')$ due to the Green--Schwarz anomaly cancellation) couple the geometry, $H$-flux and gauge bundle in an intricate way. This implies that the gauge and geometric moduli of heterotic compactifications can not be meaningfully separated, even infinitesimally, and indicates a rich structure in moduli space that is of pivotal importance for the understanding of the effective physical theory on space-time. These subtleties suggest that  a more complete theoretical understanding of the moduli space of heterotic systems is required.

The study of such coupled geometries is also very interesting from a mathematical perspective. Traditionally, when investigating bundles in various dimensions it is common to fix a background structure, such as Calabi--Yau, $G_2$, etc. One then studies instanton bundles over these geometries, and their corresponding classical and quantum moduli spaces. This often allows for the definition of geometric invariants, such as Donaldson--Thomas invariants in the case of holomorphic bundles (or more generally sheaves) over Calabi--Yau varieties \cite{thomas1997gauge, donaldson1998gauge, Thomas:1998uj}. It is known, however, that the ``invariants'' defined in this way tend to have residual dependence on the background structure. A simple example of such an invariant is the dimension of the moduli space of a holomorphic bundle, which in general experiences jumps across the complex structure moduli space of the background Calabi--Yau variety. In general, more complicated wall-crossing phenomena can appear and it is even debated if meaningful invariants can be defined in the case of $G_2$ structures \cite{donaldson2009gauge, Joyce:2016fij}. In light of such issues, it appears more natural to study the coupled geometric system of base and bundle from the start, thereby investigating the corresponding coupled moduli space to see if any invariants can be defined for the total geometry. The heterotic system is interesting in this regard: although the Green--Schwarz anomaly cancellation appears complicated at first sight, from a certain perspective it is actually rather natural. This approach has already proven useful for heterotic $SU(3)$ structures \cite{Anderson:2014xha,delaOssa:2014msa,Garcia-Fernandez:2015hja,Candelas:2016usb,Garcia-Fernandez:2018emx,Ashmore:2018ybe,Garcia-Fernandez:2018ypt,Candelas:2018lib} and $G_2$ structures \cite{delaOssa:2016ivz, delaOssa:2017pqy}.  The current paper fits well within this program.

In compactifications to four dimensions, $\cN=1$ supersymmetric compactifications of the heterotic string are given by the Hull--Strominger system \cite{Strominger:1986uh,Hull:1986kz}. In these constructions, supersymmetry and maximal symmetry  constrain the four dimensional space time to be Minkowski. Furthermore, the internal 6-manifold must have an $SU(3)$ structure, where the $H$-flux specifies the torsion, and the gauge bundle must satisfy an $SU(3)$ instanton condition. When there is no $H$-flux, this reduces to the Calabi--Yau compactification of Ref.~\cite{Candelas:1985en}.  It can be shown that part of the supersymmetry constraints on the internal geometry and bundle, specifically the F-term conditions, give rise to a certain nilpotent operator $\bar D$ on an extension bundle $\cal Q$. This perspective on the Hull--Strominger system  is particularly useful when determining the infinitesimal moduli of the system: they are captured by classes in the first cohomologies of the nilpotent operator, $H^1_{\bar D} ({\cal Q})$ \cite{Anderson:2014xha,delaOssa:2014msa,Garcia-Fernandez:2015hja}.\footnote{The reader is referred to Refs.~\cite{Anderson:2009nt,Anderson:2011cza,Anderson:2010mh,Anderson:2011ty} and Refs.~\cite{Kreuzer:2010ph,Melnikov:2011ez,Bertolini:2014dia,Bertolini:2017lcz} for earlier work on the infinitesimal moduli of Hull--Strominger systems.} It is noteworthy how  this  structure of the  moduli space mimics Kodaira--Spencer theory for deformations of a complex manifold.

In this paper, we study $\cN=1$ supersymmetric compactifications of the heterotic string to three dimensions in the supergravity limit. In this case, supersymmetry and maximal symmetry constrain the three dimensional geometry to be Anti de Sitter (AdS$_3$) or Minkowski, and the three dimensional cosmological constant is determined by a certain component of the $H$-flux. The simplest solutions of this type arise from compactifications on $G_2$ manifolds with an instanton bundle. In general, the internal 7-manifold must admit a $G_2$ structure, whose torsion is again specified by the $H$-flux, and the gauge bundle must satisfy a $G_2$ instanton condition  \cite{Gunaydin:1995ku,FriedrichIvanov2001,Firedrich:2003,Gauntlett:2003cy,Ivanov:2003nd}, see also \cite{Lukas:2010mf,Gray:2012md,delaOssa:2014lma}. We will refer to such configurations as $\cN=1$ heterotic $G_2$ systems \cite{delaOssa:2017pqy}. Again, there is a corresponding nilpotent operator $\check{\cal D}$ on an associated bundle $\cal Q$ (which is, however, not an extension bundle), and the infinitesimal deformations of heterotic $G_2$ systems are captured by $H^1_{\check{\cal D}} ({\cal Q})$ \cite{delaOssa:2014lma,delaOssa:2016ivz,delaOssa:2017gjq,delaOssa:2017pqy} (see also \cite{Clarke:2016qtg} for a slightly different approach). Note that $\check{\cal D}$ being nilpotent is equivalent to a slight generalisation of the $\cN=1$ heterotic $G_2$ system and this is simply referred to as the heterotic $G_2$ system in \cite{delaOssa:2017pqy}.

To obtain further information about heterotic moduli spaces, and in particular to go beyond the infinitesimal limit, it is useful to determine and study the superpotential of the effective lower dimensional theory. In the four dimensional case, this superpotential has been determined in Refs.~\cite{Gurrieri:2004dt,delaOssa:2015maa,McOrist:2016cfl}.
 As required by $\cN =1$ supersymmetry in four dimensions, it is a holomorphic functional on the off-shell parameter space of the system, and its critical locus corresponds to the infinitesimal moduli space. Adopting this perspective leads to an understanding of the structure of the finite deformations of these configurations, and the holomorphicity of the superpotential  provides constraints enough to determine a third-order Maurer--Cartan equation for finite moduli  \cite{Ashmore:2018ybe}.  
 
In this paper, we take the first step in a similar analysis of $\cN=1$ heterotic $G_2$ systems by determining the associated superpotential. A general feature of supergravity theories is a relation between  gravitino mass and the superpotential. Specifically, we have
\[
  M_{3/2}=e^{K/2}W\,,
\]
so if we know any two out of three of the mass $M_{3/2}$, superpotential $W$ and Hessian potential $K$, we can deduce the third. The gravitino mass can be computed by a reasonably straightforward dimensional reduction of the ten dimensional effective action and based on the resulting form, we propose that the superpotential is given by
\[
W = \frac{1}{4}\, \int_Y e^{-2\phi}\ \left((H + h\, \varphi)\wedge\psi 
- \frac{1}{2}\, \dd\varphi\wedge\varphi\right)~,
\]
where $H$ is the $H$-flux on the internal 7-manifold $Y$,  $h$ is a constant which determines the AdS$_3$ curvature scale, $\phi$ is the dilaton, and $\varphi, \psi$ determine the $G_2$ structure. We provide additional evidence for this superpotential by showing that it reproduces all supersymmetry constraints of $\cN=1$ heterotic $G_2$ systems at its critical locus.
Computing its second order variations will therefore reproduce the constraints for infinitesimal moduli.  We also comment on the relation between this superpotential and the Hitchin functional in the text.

 We thus expect the superpotential  $W$ to capture the low-energy physics  of the supergravity regime of the heterotic string. However, $\cN=1$ supersymmetry in three dimensions is less constraining than in four dimensions. Notably, the three dimensional superpotential is necessarily real, so we lack the holomorphicity that powers the usual nonrenormalisation theorems. Consequently, in three-dimensional supergravity there is less control of quantum corrections. We will comment more on this in the conclusions of this paper.  Presently, we content ourselves to work at leading order in $\alpha'$ and, correspondingly, restrict to the large volume, weak coupling limit where supergravity should provide a good description of the physics.

In future works, we hope to use the superpotential $W$ as a tool to better understand the finite-order moduli space and the resulting physical constraints. While the infinitesimal moduli space of these systems have been studied in recent years \cite{delaOssa:2014lma,delaOssa:2016ivz,delaOssa:2017gjq,delaOssa:2017pqy,Clarke:2016qtg}, the finite deformations are still poorly understood. One reason for this is that the tools of complex geometry, which proved so useful in the four dimensional case,  rely on holomorphicity of the superpotential, a property we lose in three dimensions. We intend to come back to this question in future publications.

The basic mathematical setting of this paper is the moduli space of heterotic systems. We begin in Section \ref{sec:bground} by briefly reviewing the inherent geometry and the constraints that string theory puts on them. In Section \ref{sec:dimredux} we compute the mass functional of the three dimensional gravitino and extract from this a candidate superpotential for the effective physics. In Section \ref{sec:supotvar} it is shown that the critical locus of this functional does in fact reproduce the known supersymmetry constraints, providing further evidence that it plays the part of a superpotential. We conclude in Section \ref{sec:conc} and give some indications of future applications. In order to correctly identify the superpotential, it was important to correctly normalise our fields and, since we include external space flux, to identify the physical mass in AdS$_3$ space. These technical calculations are recorded in Appendix \ref{ap:conv}, which also summarises our conventions.

\section{Background}\label{sec:bground}
\subsection{$G_2$ structures}\label{ssec:g2struct}

A seven dimensional manifold is said to have a $G_2$ structure if it admits a non-degenerate associative 3-form  $\varphi$. Such a $G_2$ structure exists when the first and second Stiefel-Whitney classes of $Y$ are trivial, that is when $Y$ is orientable and spin.  The form $\varphi$ determines a Riemannian metric $g_\varphi$
on $Y$, and a coassociative 4-form $\psi = *\varphi$, where the Hodge dual is taken using the $G_2$ metric $g_\varphi$. We refer the reader to \cite{MR916718,bonan66,FerGray82,Hitchin:2000jd,joyce2000,Bryant:2005mz,delaOssa:2016ivz} for more details on $G_2$ structures.

The exterior derivative of the forms $(\varphi,\psi)$, which give the structure equations for the $G_2$ structure, can be decomposed into irreducible representations of $G_2$  
\begin{align*}
\dd_7\varphi &= \tau_0 \, \psi + 3\, \tau_1\wedge\varphi + *\tau_3
~,
\\
\dd_7\psi&= 4\, \tau_1\wedge\psi + *\tau_2~,
\end{align*}
where the torsion classes $\tau_k$ are $k$ forms,
$\tau_3$ is in the $\bf 27$ irreducible representation of $G_2$ and
$\tau_2$ in the $\bf 14$  \cite{Gauntlett:2001ur,Firedrich:2003,FriedrichIvanov2001,Lukas:2010mf,Gray:2012md,delaOssa:2014lma}. 
In the case that the $G_2$ structure on $Y$ is integrable (that is, $\tau_2 = 0$) \cite{fernandez1998dolbeault} the structure equations can be written
\begin{align}
\dd_7\varphi &= \tau_0 \, \psi + 3\, \tau_1\wedge\varphi + *\tau_3
= i_T(\varphi)~,\label{eq:g2phi}
\\
\dd_7\psi&= 4\, \tau_1\wedge\psi = i_T(\psi)
~,\label{eq:g2psi}
\end{align}
for
\be
T(\varphi)= \frac{1}{6}\,\tau_0\, \varphi - \tau_1\lrcorner_7\, \psi - \tau_3~.\label{eq:torsion}
\ee
We remark that a $G_2$ structure admits a totally antisymmetric torsion if and only if 
 $\tau_2 = 0$, and $T(\varphi)$ is in fact the torsion of the unique metric connection $\Dtor$ with totally antisymmetric torsion \cite{FriedrichIvanov2001,Bryant:2005mz,delaOssa:2017pqy}.

Alternatively, we may discuss $G_2$ structures in terms of spinors. Indeed, a seven dimensional manifold has $G_2$ structure if it admits a nowhere vanishing spinor $\lambda$. The spinor, $\lambda$, in fact defines a globally-defined, nowhere-vanishing three-form with components:
  \begin{align}
    -i \lambda^\dagger\Gamma_{ijk}\,\lambda &=\varphi_{ijk}\,,\label{G27}
  \end{align}
  which is identified with the associative three-form discussed above. Furthermore, $\lambda$ satisfies the following differential equation
  \begin{align}
   \nabla^{G_2}_i\,\lambda = \DD_i\lambda
   -\frac{1}{8}\,T_{ijk}\,\Gamma^{jk}\,\lambda&=0\label{eq:KS7}
  \end{align}
where $D$ is the Levi-Civita connection and $T$ is identified with the torsion of the $G_2$ structure. 

\subsection{Heterotic $G_2$ systems} \label{ssec:hetg2}
Let $Y$ be a seven dimensional manifold with a $G_2$ structure $\varphi$ and let $V$ be a vector bundle on $Y$ with connection $A$.   We are interested in compactifications of the ten dimensional heterotic superstring on $(Y, V)$ which preserve minimal supersymmetry.   This requirement constrains the allowed geometry of the compactifications  \cite{Gunaydin:1995ku,FriedrichIvanov2001,Firedrich:2003,Gauntlett:2003cy,Ivanov:2003nd,Lukas:2010mf,Gray:2012md,delaOssa:2014lma}.  We call the resulting geometry an $\cN=1$ heterotic $G_2$ system.   A heterotic $G_2$ system \cite{delaOssa:2017pqy} is defined to be the quadruple
\be
[ (Y, \varphi), (V, A), (TY, \Theta), H]~,
\ee
where
\begin{itemize}
\item $\varphi$ is an integrable $G_2$ structure on the seven dimensional manifold $Y$ (see section \ref{ssec:g2struct} for definitions).
\item $V$ is a gauge bundle with connection $A$ whose curvature $F$ satisfies an instanton condition
\be
F\wedge\psi = 0~.\label{eq:Vinst}
\ee

\item $\Theta$ is a connection on the tangent bundle $TY$ of $Y$ which is also an instanton
\be
R(\Theta)\wedge\psi = 0~,\label{eq:TYinst}
\ee
where $R(\Theta)$ is the curvature of $\Theta$.
\item $H$ is a three form defined by 
\be
H = \dd B + \frac{\alpha'}{4}\, ({\cal CS}(A) - {\cal CS}(\Theta))~,\label{eq:flux}
\ee
where ${\cal CS}(A)$ is the Chern--Simons form of the connection $A$
\be
{\cal CS}(A) = \tr \left( A\wedge \dd A + \frac{2}{3}\, A^3\right)~,\label{eq:anom}
\ee
with a similar definition for ${\cal CS}(\Theta)$, and  $B$ is the $B$-field.
The fields $H$, $A$, $B$ and $\Theta$ are constrained such that
\be
H = T(\varphi)~,\label{eq:anomaly}
\ee
where $T(\varphi)$ is given in equation \eqref{eq:torsion} and it is the totally antisymmetric torsion of a (unique) connection $\Dtor$ compatible with the $G_2$ structure.
\end{itemize}
Finally, $\cN =1$ supersymmetry imposes a relation between the dilaton and the torsion, namely 
\[ \tau_1 = \frac{1}{2} \dd \phi \, ,\]
and that the external part of the flux is proportional to the scalar part of the $G_2$ torsion:
\[
 h = \frac{1}{3} \tau_0 \, ,
\]
where $h$ will be defined in terms of the external flux below. We will see below that $h$ is constant, i.e. coordinate independent.

\subsection{The heterotic $G_2$ system as a differential}\label{sec:G2diff}

It was noticed in \cite{delaOssa:2017pqy} that the heterotic $G_2$ system is equivalent to the existence of a differential complex $(\check\Omega^*({\cal Q}),\check{\cal D})$, where $\check{\cal D}$ is the corresponding differential, and $\check\Omega^*({\cal Q})$ is a sub-complex of $\Omega^*({\cal Q})$. In this subsection, we will recapitulate the definitions of $\check{\cal D}$ and $\check\Omega^*({\cal Q})$, referring to \cite{delaOssa:2017pqy} for more details and relevant references on this topic.

Given a manifold $(Y,\varphi)$ together with a vector bundle ${\cal G}={\rm End}(V)\oplus{\rm End}(TY)$, where ${\rm End}(V)$ has structure group contained in $E_8\times E_8$, one can construct the bundle ${\cal Q}=TY\oplus{\cal G}$ and consider the complex $\check\Omega^*({\cal Q})\subseteq \Omega^*({\cal Q})$ given by projecting two-forms to the $\bf 7$ representation, three-forms to the singlet representation, and higher degree forms to zero. Using the $G_2$ structure and vector bundle data, one can construct an operator 
\begin{equation}
\check{\cal D}:\;\;\;\check\Omega^p({\cal Q})\;\;\longrightarrow\;\;\check\Omega^{p+1}({\cal Q})\:, 
\end{equation}
given by $\check{\cal D}=\pi\circ{\cal D}$, where $\pi$ denotes the appropriate projection, and ${\cal D}$ is a connection on $\Omega^*({\cal Q})$ which we will now define. 

Explicitly, we have
\begin{equation}
{\cal D}=\dd+\zeta+{\IA}+{\cal F}\:.
\end{equation}
Here,  $\IA= A + \Theta$ is the connection on $\cal G$ and  $\zeta$ is and connection on $TY$ given by
\begin{equation}
\zeta^n{}_m = \Gamma^n{}_{mp}\,\dd y^p\:,
\end{equation}
where $\Gamma^n{}_{mp}$ are the connection symbols of the $BPS$ connection $\Dtor$, i.e. the connection preserving the $G_2$ spinor. Note that the connection symbols in $\Dtor$ and those corresponding to  $\dd_{\zeta}=\dd+\zeta$ are not the same except when $\Dtor$ is the Levi-Civita connection. In fact, the torsion of the connection $\dd_\zeta$ is $-T$. 

The map ${\cal F}$ is constructed using the curvature $\IF$ of $\IA$. It acts on forms with values in $TY$ or $\cal G$. For $y\in\Omega^p(Y, TY)$, and $\alpha\in \Omega^p(Y,{\cal G})$ we have 
 \begin{align*}
 {\cal F} :\quad \Omega^p(Y, TY)\oplus \Omega^p(Y,{\cal G})
 &\longrightarrow 
 \Omega^{p+1}(Y,{\cal G})\oplus \Omega^{p+1}(Y, TY)
 \\
 \left(\begin{matrix}
 y  \\ \alpha 
 \end{matrix}\right)\qquad\qquad\quad
 &\mapsto
 \quad\qquad \qquad  
\left(\begin{matrix}
 {\cal F}(y) \\  {\cal F}(\alpha) 
 \end{matrix}\right)\qquad
  \end{align*}
  where
  \begin{align*}
  {\cal F}(y) &= (-1)^{p}\, y^b\wedge F_{bc}\, \dd x^c~,
  \\
  {\cal F}(\alpha)^a &= (-1)^{p}~ \frac{\alpha'}{4}\, \tr (\alpha\wedge {F^a}_b\, \dd x^b)~.
  \end{align*}
It was then shown in \cite{delaOssa:2017pqy} that the operator $\check{\cal D}$ is nilpotent, i.e. $\check{\cal D}^2=0$ and the operator $\check{\cal D}$ is a differential if and only if the $G_2$ structure and bundles satisfy the heterotic $G_2$ system, {\it including} the rather complicated heterotic Bianchi identity. Note in particular that this implies that $\cal D$ is itself an instanton connection on ${\cal Q}$. Furthermore, the infinitesimal moduli of heterotic $G_2$ systems were also shown to correspond to classes in  $H^1_{\check{\cal D}}({\cal Q})$. We will start to connect some threads between this moduli analysis and the superpotential considerations of the current paper in the outlook part of section \ref{sec:conc}.

\section{Dimensional Reduction} \label{sec:dimredux}

Let $ [ (Y, \varphi), (V, A), (TY, \Theta), H]$ be a heterotic $G_2$ system with exact $\tau_1$, as defined in section \ref{ssec:hetg2}. Compactification of the heterotic string on this system leads to a minimally  supersymmetric ($\cN=1$) effective field theory on either AdS$_3$  or Minkowski space time.

In this section we use dimensional reduction to determine the three dimensional superpotential arising from compactifications of the heterotic string on manifolds of $G_2$ structure. Our analysis follows the logic of \cite{Gurrieri:2004dt}: we will first determine all contributions to the three dimensional gravitino mass $M_{3/2}$, and then read off the superpotential from the relation
\be
M_{3/2} = e^K W \; .
\ee

The contributions to the three dimensional gravitino mass arise from the fermionic part of the ten dimensional action of heterotic supergravity
 \cite{GSW2,Bergshoeff1989439,Gurrieri:2004dt}
\begin{equation} 
\begin{split}
    \label{eq:faction} S_{0,f}=-\frac{1}{2 \kappa_{10}^2}&\int_{M^{10}}d^{10}x\sqrt{-g}\,e^{-2\phi} \\
    &\left(\overline{\Psi}_M\Gamma^{MNP} D_N\Psi_P-\frac{1}{24}\left(\overline{\Psi}_M\Gamma^{MNPQR}\Psi_R+6\,\overline{\Psi}^N\Gamma^P\Psi^Q\right)H_{NPQ}\right)\, .
    \end{split}
  \end{equation}
The three dimensional action of the gravitino contains kinetic  and mass terms
   \begin{equation}\label{eq:RSeqn}
    -\frac{1}{2 \kappa_3^2}\int d^3x\sqrt{-g}\left(\overline{\psi}_\mu\Gamma^{\mu\nu\kappa} D_\nu\psi_\kappa+m\overline{\psi}_\mu\Gamma^{\mu\kappa}\psi_\kappa\right)\,,
  \end{equation}
and we can identify contributions to the mass from terms in the ten dimensional action that have two three dimensional Clifford matrices contracted with three dimensional gravitinos.  Finding such terms is straightforward, as is the dimensional reduction once we have settled the normalisation of our fields. In particular, we must ensure that the dimensional reduction results in a canonical Einstein--Hilbert term and define the gravitino mass in AdS$_3$ space with care.

\subsection{Gravitino mass terms} 

In dimensionally reducing the fermionic action \eqref{eq:faction}, all three terms contribute to the three dimensional gravitino mass term. In this section, we perform the dimensional reduction and rewrite the respective contributions in a language adapted to the $G_2$ structure of the compact 7-manifold.\footnote{We will set $\kappa_{10}=1$ for this calculation, reinstating  factors of $\kappa_{10}$ in the next subsection.} Some elements of the calculation and, in particular, our notational conventions  are relegated to appendix \ref{ap:conv}.

 \subsubsection*{Mass term 1 - three $\Gamma$-term}
 
The ten dimensional gravitino kinetic term contributes to the three dimensional mass by taking the covariant derivative along the internal space directions, i.e.
 \begin{align*}
  &\int d^{10}X\sqrt{-g_{10}}\,e^{-2\phi}\overline{\Psi}_\mu\Gamma^{\mu i\nu}\DD_i\Psi_\nu\\
  =&\int d^3xd^7y\sqrt{-g_3}\sqrt{g_7}\,e^{-2\phi+3n/2+2\beta+3\alpha} \times\\
  &\quad\times(\bar{\rho}_\mu\otimes\lambda^\dagger\otimes\theta^\dagger\sigma^2)(\Gamma^{\nu\mu}\otimes\Id\otimes(\sigma^2)^2)(\Id\otimes\Gamma^i\otimes\sigma^1)\DD_i(\rho_\nu\otimes\lambda\otimes\theta)\\ 
 =&\int d^3x \sqrt{-g_3}\,(\bar{\rho}_\mu\Gamma^{\mu\nu}\rho_{\nu})\left[-\int d^7y\sqrt{g_7}\, e^{-2\phi+3n/2+2\beta+3\alpha} (-i\lambda^\dagger\Gamma^i\DD_i\lambda) \right]\,.
 \end{align*}
 In the last line, we find a quadratic term for the three dimensional gravitino and hence interpret the expression in the square bracket as mass term.  In the second line\footnote{A priori, we should also include a term from the derivative of the scale factor $\beta$, proportional to $(\DD_i\beta)\lambda^\dagger\Gamma^i\lambda$; however, $\lambda^\dagger\Gamma^i\lambda$ corresponds to a $G_2$-invariant vector and must therefore vanish. } we have used the decomposition of the ten dimensional metric, spinor and $\Gamma$ matrices discussed in appendix \ref{ap:conv}. In particular, motivated by considerations of the canonical three dimensional Einstein--Hilbert term and gravitino kinetic term, we include scale factors (summarised in \eqref{eq:metcons}--\eqref{eq:fermcons2})  when decomposing the ten dimensional fields. The importance of these factors will be discussed in more detail below.

We now recall that $\lambda$ satisfies the seven dimensional Killing spinor equation \eqref{eq:KS7}. Consequently, $\lambda$ determines a $G_2$-structure with positive $G_2$ three-form $\varphi$ given by \eqref{G27}. Using this, we can rewrite the mass contribution as
\begin{align}
  M_1=-\frac{1}{8}\int d^7y\sqrt{g_7}\,T_{ijk}\,\varphi^{ijk} \,e^{-2\phi+3n/2+2\beta+3\alpha}\,.
\end{align}
where $T$ is the torsion of the $G_2$ structure. It is therefore related to the exterior derivative of $\varphi$:
\begin{align}
  d\varphi=\frac{1}{4}T_{ij}^{\phantom{as}e}\varphi_{ekl}\,dx^{ijkl}\,,
\end{align}
which implies, after using an identity from \cite[App.A]{delaOssa:2017pqy}, that 
\[ T_{ijk}\,\varphi^{ijk}=d\varphi\lrcorner\psi=*_7(d\varphi\w\varphi)~.\]
 In particular, this determines the contribution to the gravitino mass as
\begin{align}
  M_1=-\frac{1}{8}\int d\varphi\w\varphi\cdot e^{-2\phi+3n/2+2\beta+3\alpha}\,.\label{eq:masscon1}
\end{align}

\subsubsection*{Mass term 2 - five $\Gamma$-term}

The second term in the fermionic action \eqref{eq:faction} contributes to the gravitino mass when both ten dimensional gravitino indices are along the three dimensional spacetime and the remaining three $\Gamma$-matrices have indices along the internal space. That is:
\begin{align*}
& \int d^{10}X\sqrt{-g_{10}}\,e^{-2\phi}\left(-\frac{1}{24}\overline{\Psi}_\mu\Gamma^{\nu\mu}\Gamma^{ ijk}\Psi_\nu H_{ijk} \right)\\
  =&\int d^3x d^7y\,\sqrt{-g_3}\sqrt{g_7}\,e^{-2\phi+3n/2+2\beta+3\alpha}\times\\&\quad\times
 \left(-\frac{1}{24}\right) (\bar{\rho}_\mu\otimes\lambda^\dagger\otimes\theta^\dagger\sigma^2)(\Gamma^{\nu\mu}\otimes\Id\otimes\Id)(\Id\otimes\Gamma^{ijk}\otimes\sigma^1)(\rho_\nu\otimes\lambda\otimes\theta)H_{ijk}\\
  =&\int d^3x\sqrt{-g_3}\,(\bar{\rho}_\mu\Gamma^{\mu\nu}\rho_\nu)\left[ \frac{1}{24}\int d^7y\sqrt{g_7}\,e^{-2\phi+3n/2+2\beta+3\alpha}
  (-i\lambda^\dagger\Gamma^{ijk}\lambda)H_{ijk}\right]\,.
\end{align*}
where the computation again relies on the conventions discussed in appendix \ref{ap:conv}. Recognizing the $G_2$ three-form in the spinor bilinear ({\it cf.}~Eq.~\eqref{G27}), the mass-contribution in the last row contains $\varphi^{ijk}H_{ijk}$, which we can rewrite as $6*_7(H\w\psi)$. Consequently, the mass contribution is
\begin{align}
  \label{eq:masscon2}
  M_2&=\frac{1}{4}\int *_7H\w\varphi \cdot e^{-2\phi+3n/2+2\beta+3\alpha}\,.
\end{align}

\subsubsection*{Mass term 3 - single $\Gamma$-term}
 
The final contribution to the three dimensional gravitino mass originates from the single $\Gamma$-term in \eqref{eq:faction}, namely
\begin{align}
  \label{eq:10d1gamtermA}
  \int d^{10}X\sqrt{-g_{10}}\,\left( -\frac{1}{4}e^{-2\phi}\,\overline{\Psi}^N\Gamma^P\Psi^QH_{NPQ} \right) \,.
\end{align}
When both ten dimensional gravitino indices are along the three dimensional spacetime, dimensional reduction  gives
\begin{align*}
  &\int d^{10}X\sqrt{-g_{10}}\left( -\frac{1}{4} e^{-2\phi}\,\overline{\Psi}_{\xi}\Gamma^\sigma\Psi_{\omega} \,H_{\kappa\sigma\lambda}\,g^{\kappa\xi}_{(10)}g^{\lambda\omega}_{(10)} \right)\\
  =&\int d^3x\, d^7y\,\sqrt{-g_3}\sqrt{g_7}\,e^{-2\phi+3n/2-2n+2\alpha+2\beta} \times\\ &\quad\times
 \left( -\frac{1}{4} \right) (\bar{\rho}_\xi\otimes\lambda^\dagger\otimes\theta^\dagger\sigma^2)(\Gamma^\sigma\otimes\Id\otimes\sigma^2)(\rho_\omega\otimes\lambda\otimes\theta) H_{\kappa\sigma\lambda}\,g^{\kappa\xi}_{3}g^{\lambda\omega}_3\\
  =&\int d^3xd^7y\,\sqrt{-g_3}\sqrt{g_7}\,e^{-2\phi+3n/2-2n+2\alpha+2\beta} (\bar{\rho}_\xi\Gamma^\sigma\rho_\omega)
  \left( -\frac{1}{4} \right) H_{\kappa\sigma\lambda}\,g^{\kappa\xi}_3g^{\lambda\omega}_3\,.
\end{align*}
While this is a quadratic term for the three dimensional gravitino, it does not yet have the right $\Gamma$ matrix structure to be interpreted as a mass term. However, in three dimensions the  Clifford duality \cite{Polchinski:1998rr}
\begin{align}
  \Gamma^\sigma&=-\frac{1}{2}\epsilon^{\sigma\mu\nu}\Gamma_{\mu\nu}\sqrt{-g_3}\,,
\end{align}
can be inserted in the dimensionally reduced action term, with result
\begin{align*}
  &\int d^3xd^7y\,\sqrt{-g_3}\,\sqrt{g_7}\,e^{-2\phi+3n/2-2n+2\alpha+2\beta}(\bar{\rho}_\xi\Gamma_{\mu\nu}\rho_\omega)
  \left[\frac{1}{8} \epsilon^{\sigma\mu\nu}H_{\sigma\lambda\kappa}\sqrt{-g_3}\,g^{\kappa\xi}_3g^{\lambda\omega}_3\right]\,.
\end{align*}

Clearly, the  three-form flux along the three dimensional spacetime will be proportional to the totally antisymmetric form, $H_{\sigma\lambda\kappa}=\hM\epsilon_{\sigma\lambda\kappa}$ for some $\hM$. 
Moreover, $\hM$ can be shown to be related to $*_3H_{(3)}$:
\[
 *_3H_{(3)}=\frac{\sqrt{- g_3}}{3!}\,\hM\,\epsilon^{\mu\nu\sigma}\,\epsilon_{\mu\nu\sigma}=\frac{\hM}{\sqrt{-g_3}}\,.
\]
We may use this to write the gravitino mass term in a succinct form
\begin{align*}
 &\int d^3x d^7y\,\sqrt{-g_3} \sqrt{g_7}\,e^{-2\phi+3n/2-2n+2\alpha+2\beta}(\bar{\rho}_\xi\Gamma_{\mu\nu}\rho_\omega)
    \left[\frac{1}{8} *_3H_{(3)}\,g_3^{\kappa\xi}g_3^{\lambda\omega}\,(\delta^\mu_\lambda\delta^\nu_\kappa-\delta^\mu_\kappa\delta^\nu_\lambda) \right] \\
  =&\int d^3x\sqrt{-g_3}\,\bar{\rho}_\mu\Gamma^{\mu\nu}\rho_{\nu}
  \left[-\frac{1}{4}\int d^7y\sqrt{g_7}\,*_3H_{(3)}\,e^{-2\phi+3n/2-2n+2\alpha+2\beta}\right]\,.
\end{align*}
Finally, the mass contribution is simplified if rewritten  in terms of the flux parameter $f = e^{-3n/2} *_3H_{(3)}$ ({\it cf.}  \eqref{eq:fstarH}) 
\begin{align}
  \label{eq:masscon32}
  T_3&=-\frac{1}{4} \int d^7y\sqrt{g_7}\, f\, e^{-2\phi+n+2\alpha+2\beta}\,.
\end{align}

\subsubsection*{Total mass contribution}

We now simply collect together the three mass contributions \eqref{eq:masscon1}, \eqref{eq:masscon2} and \eqref{eq:masscon32}:
\be
  \tilde{M}_{3/2}=-\frac{1}{8}\int_7 e^{-2\phi + n}\left(d\varphi\w\varphi-2*_7H\w\varphi+2*_7f\right)\, ,
\ee
after we have used that consistency of the Clifford algebra imposes that the scale factors are related by $\beta=-\alpha=\frac{n}{2}$.

This is not, however, the mass contribution we would like to use since we have to account for the corrected kinetic term in AdS$_3$ space discussed in appendix \ref{sec:apAdSKin}. Indeed, it is possible to use conventions for the gravitino mass so that supersymmetric solutions have a massless gravitino both in Minkowski and AdS$_3$ space, and we find these conventions useful for the analysis performed in this paper. As shown in the appendix, this can be accomplished using a three dimensional covariant derivative $\nabla_{\mu}$, which is shifted by a term proportional to the cosmological constant ({\it cf.}~Eq. \eqref{eq:apAdSCovDer}). Rewriting the gravitino kinetic term  in terms of this operator, we must shift the mass term accordingly ({\it cf.}~Eq \eqref{eq:3dkinAdS}). The result is that the gravitino is governed by the action 
  \begin{equation}\label{eq:RSeqn}
    \int d^3x\,\sqrt{-g}\left(\overline{\rho}_\mu\Gamma^{\mu\nu\kappa}\nabla_\nu\rho_\kappa+{M}_{3/2}\,\overline{\rho}_\mu\Gamma^{\mu\kappa}\rho_\kappa\right)\,,
  \end{equation}
  where
  \begin{align}
 M_{3/2}&=-\frac{1}{8}\int_7 e^{-2\phi +n}\left(d\varphi\w\varphi-2*_7H\w\varphi+4*_7f\right)\,.
\end{align}
This expression simplifies somewhat if we introduce the flux parameter $h=-\frac{2}{7} f$, whence
 \begin{align}
 M_{3/2}&=\frac{1}{4}\int_7 e^{-2\phi +n}\left(-\frac{1}{2} d\varphi\w\varphi+(H + h \varphi)\w\psi\right)\,.
\end{align}

\subsection{Three dimensional Einstein--Hilbert term}  \label{sec:3dEHterm}

As discussed above we require that the three dimensional $\cN=1$ theory has a canonical Einstein--Hilbert term 
\begin{align}
  \frac{1}{\kappa_3^2}\int_A d^3x\sqrt{-g_3}\,R_3\,.
\end{align}
In general, this necessitates a conformal rescaling of the ten dimensional metric as in (\ref{eq:metcons}). We here fix this factor. 

Under a conformal transformation $g\rightarrow e^sg$, the Ricci scalar transforms as $R\rightarrow e^{-s}R+\ldots$, where the ellipsis indicates irrelevant terms, not proportional to $R$. Using this fact and equation (\ref{eq:metcons}), the dimensional reduction is:
\begin{align}
 -\frac{1}{2\kappa^2_{10}} \int d^{10}X\sqrt{-g_{10}}\,e^{-2\phi} R&\rightarrow -\frac{1}{2\kappa^2_{10}}\int d^{10}X\sqrt{-g_3}\sqrt{g_7}\, e^{-2\phi+3n/2 } R_3\,e^{-n}\nonumber\\
  =-&\frac{1}{2\kappa^2_{10}}\int d^3x\sqrt{-g_3}\,R_3 \left(\int d^7y\sqrt{g_7}\,e^{-2\phi+n/2}\right)\,.
\end{align}
We recall that the dilaton $\phi$ may depend on the coordinates of the $G_2$ structure manifold. However, three dimensional Lorentz invariance forbids a dependence on the non-compact dimensions, and so $\phi = \phi(y)$. We can then define the constant volume scale
\begin{align} \label{eq:ehscale}
v=\int d^7y\sqrt{g_7}\,(e^{-2\phi+n/2})
\end{align}
Setting $\kappa_3^2=\kappa_{10}^2/v$, we then recover the canonically-normalised Einstein--Hilbert term
\be
\frac{1}{2\kappa^2_{3}}\int d^3x\sqrt{-g_3}\,R_3 \,.
\ee
Note that this constrains $n$ to be a function of $y$, independent of the non-compact coordinates. In fact, $n$ has to be constant in the sense that it does not depend on the internal coordinates either. Indeed, as we will show in section \ref{sec:supotvar} the seven dimensional BPS equations can be derived from setting the superpotential and it's variation to zero, or equivalently the gravitino mass and it's variation to zero. It turns out that if $n$ is not constant internally, then this changes the $\bf 7$ part of the torsion and we do not get agreement with the known BPS equation. Hence $n$ must be constant internally as well.


Indeed, in the three dimensional $\cN=1$ theory we are striving to determine, the dynamical fields are given by fluctuations of the ten dimensional fields around the vacuum solution. Let us focus on the two fluctuations that can change $v$, namely the fluctuations  $(\delta{\phi}, \delta V)$ about the vacuum expectation values of, respectively, the dilaton $\phi$ and internal volume 
\[
V= \int d^7y\sqrt{g_7}\, .
\]
Under such fluctuations, $v$, defined by \eqref{eq:ehscale}, can only stay constant  if  
\[
n \to n + \delta n = n + 4 \delta{\phi}-2\ln \delta V \; .
\]
Thus, fluctuations of $n$ are essential for the three dimensional theory  to have a canonical gravitational sector. This determines how $n$ depends on the parameters.

Finally, let us remark that the required variation of $n$ nicely matches the expected form of a Hessian potential for the metric on the part of the heterotic $G_2$ parameter space spanned by $(\delta{\phi}, \delta V)$. With this match in mind, we will from now on assume that $n$ is constant in vacuum, so that \be \label{eq:Kn}
K\simeq n \; ,
\ee
is a reasonable identification. Furthermore, by the ten dimensional Bianchi identity it follows that $\tilde h$ is constant on the internal manifold. From this we can also deduce that $f$ must be constant.

\subsection{Relation to a three dimensional superpotential}

The above dimensional reduction fixes the gravitino mass to be
\begin{align}
 M_{3/2}&=
\,\frac{1}{4}\,e^K\int_Y e^{-2\phi}\ \left((H + h\, \varphi)\wedge\psi 
- \frac{1}{2}\, \dd\varphi\wedge\varphi\right)~.
\end{align}
In terms of three dimensional $\cN =1 $ theories, this result has the following interpretation.  With our conventions, the gravitino mass of such theories should be determined by a Hessian potential $K$ and a superpotential $W$  according to
\[
M_{3/2}=e^KW \; .
\]

  Using the above discussion of the Einstein--Hilbert term, we have tentatively identified $K \simeq n$, implying that the superpotential of the three dimensional $\cN =1 $ theory must be
\be
W = \frac{1}{4}\int_Y e^{-2\phi}\ \left((H + h\, \varphi)\wedge\psi 
- \frac{1}{2}\, \dd\varphi\wedge\varphi\right)~.
\ee
In the next section, we will provide further evidence for this conclusion.

\section{The Superpotential and the Supersymmetry Conditions} \label{sec:supotvar}

We have shown  in section \ref{sec:dimredux} that, up to an overall constant, the superpotential $W$ of the $\cN=1$ effective theory obtained by compactifying the heterotic string on a heterotic $G_2$ system $ [ (Y, \varphi), (V, A), (TY, \Theta), H]$ 
is given by
\be \label{eq:W}
W = 
\frac{1}{4} \int_Y e^{-2\phi}\ \left((H + h\, \varphi)\wedge\psi 
- \frac{1}{2}\, \dd\varphi\wedge\varphi\right)~,
\ee
where $h = -\frac{2}{7} f$   is  related to the curvature of the three dimensional spacetime by \eqref{eq:3dcc}  and $\phi$ is the dilaton field. 
In this section we show that this superpotential is a functional  whose critical points give the conditions for preservation of $\cN=1$ supersymmetry in three dimensions, or equivalently, the conditions that define the heterotic $G_2$ system discussed in section \ref{ssec:hetg2}. 
Our presentation uses machinery developed for the analysis of infinitesimal moduli of heterotic $G_2$ systems of Refs.~\cite{delaOssa:2016ivz,delaOssa:2017pqy}.

 We can view our three dimensional effective theory as a sigma model with values in some target space, locally parameterised by the scalar components of the three dimensional supermultiplets.  The superpotential controls supersymmetric vacua by requiring that its variations with respect to these scalars vanish, $\frac{\delta W}{\delta \sigma}=0$, so in order to check that our functional truly reproduces the heterotic $\cN=1$ supersymmetry conditions, thus deserving to be named superpotential, we must compute the variations with respect to the three dimensional scalars.  In the present case, these are given by $\{\phi,\varphi,\psi, B,A,\Theta\}$. Note that $h$, being (the dual of) a 3d field strength ought not be varied. 

 The variation of $\varphi$ can be written in terms of a one-form valued in $TY$,
$M^b = M_a{}^b \dd x^a \in \Omega^1(Y,TY)$ 
\be
\delta \varphi = i_M(\varphi) = \frac{3}{7}\, (\tr M) \, \varphi + i_m(\varphi )+ i_{\mathring M}(\varphi)~.
\ee
Here, in the second equality, we have defined $m$ as the two form corresponding to the antisymmetric part of $M_{ab}$, and $\mathring M$ is the traceless symmetric part of $M_{ab}$.  Note that this decomposition is precisely the partition of $\delta\varphi$ as a three form into irreducible representations of $G_2$:\footnote{Similar partitions into $G_2$ irreducible representations exist for any differential form.  } 
\be
\pi_{\bf 1}(\delta\varphi)= \frac{3}{7}\, (\tr M)  \varphi ~,\quad
\pi_{\bf 7}(\delta\varphi)= i_m(\varphi) = - (m\lrcorner\varphi)\lrcorner\psi ~,\quad
\pi_{\bf 27}(\delta\varphi)= i_{\mathring M}(\varphi)~.
\ee
Note moreover that $\pi_{\bf 14} m$ does not contribute to $\delta\varphi$ and that 
the variations corresponding to $\pi_{\bf 7} m$ leave the $G_2$ metric invariant as
\be
\delta g_{ab} = 2 M_{(ab)}~.
\ee
The variations of $\psi$ are determined by the variations of $\varphi$ and to first order
\be
\delta\psi = i_M(\psi) = \frac{4}{7}\, (\tr M) \, \psi+ i_m(\psi)
+ i_{\mathring M}(\psi)~,
\ee
where
\be
\pi_{\bf 1}(\delta\psi)= \frac{4}{7}\, (\tr M)  \varphi ~,\quad
\pi_{\bf 7}(\delta\psi)= i_m(\psi) =   (m\lrcorner\varphi)\wedge\varphi ~,\quad
\pi_{\bf 27}(\delta\psi)= i_{\mathring M}(\psi)
\ee
The Green--Schwarz anomaly cancellation mechanism implies that the variation of $H$ must be 
\be
\delta H = \dd{\cal B}+ \frac{\alpha'}{2} (  \tr( F \delta A) - \tr(R(\Theta)\,\delta\Theta))~,\label{eq:deltaH}
\ee
where, up to a closed form, we have defined 
\be
{\cal B} = \delta B - \frac{\alpha'}{4} \tr(A\,\delta A - \Theta\delta\Theta)~.\label{eq:curlyB}
\ee

The critical locus of $W$ is therefore given by
\be \label{eq:critloc}
\frac{\delta W}{\delta \Phi} = 0
\ee
where $\delta \Phi$ represents any of the variations\footnote{Note that $\dd{\cal B}$  is gauge invariant, hence $\cal B$ is invariant up to a closed form.}
\be
\delta \Phi = \{\delta\phi, \tr M, \pi_{\bf 7}(m), \mathring M, {\cal B}, \delta A, \delta\Theta\}~.
\ee

For consistency, the locus specified by \eqref{eq:critloc} should be the supersymmetric locus in the parameter space of heterotic strings. 
Consider a first order variation $\delta W$ of $W$: 
\begin{align*}
\delta W &= \int_Y e^{-2\phi} \left\{
-2\, \delta\phi \Big(
(H + h\, \varphi)\wedge\psi - \frac{1}{2}\, \dd\varphi\wedge\varphi\Big)
+ ( \delta H  + h \delta \varphi)\wedge\psi 
\right.
\\[5pt]
&\qquad  \qquad \left.
+ (H + h \varphi)\wedge\delta\psi 
- \frac{1}{2}\, \Big( \dd\delta\varphi \wedge\varphi + \dd\varphi\wedge\delta\varphi\Big)
\right\}
\\
&=  \int_Y e^{-2\phi} \left\{
-2\, \delta\phi \Big(
(H + h\, \varphi)\wedge\psi - \frac{1}{2}\, \dd\varphi\wedge\varphi\Big)
\right.
\\[3pt]
&\qquad\qquad\qquad - {\cal B}\wedge e^{2\phi} \dd (e^{-2\phi}\, \psi) 
+ \frac{\alpha'}{2}\,\left[ \tr(\delta A \, F) - \tr(\delta \Theta \, R(\Theta)) \right]\wedge\psi
+ 
\\[5pt]
&\qquad\qquad\qquad \left.
+ (H+h\varphi)\wedge\delta\psi 
+ \delta\varphi\wedge\Big( h\, \psi -\dd\varphi + \dd\phi\wedge\varphi\Big)
\right\}
\end{align*}
where, for the second equality, we have used \eqref{eq:deltaH} and \eqref{eq:curlyB}, and integrated by parts the terms containing $\dd{\cal B}$ and $\dd\delta\varphi$.  We demand that this expression should vanish for any variations of the field, and so the equation decomposes into several conditions, one for each independent field variation.

The vanishing of $\delta W$ with respect to $\cal B$ gives
\[
\dd(e^{-2\phi}\, \psi) = 0~,
\]
which is equivalent to the statement that the $G_2$ structure on $Y$ must be integrable ($\tau_2 = 0$) and that 
\be
\tau_1 = \frac{1}{2} \dd\phi~.\label{eq:pretau1}
\ee
The variations of $W$ with respect to $A$ and $\Theta$ vanish if and only if
\[
F\wedge\psi = 0 = R(\Theta) \w \psi
\]
which are the conditions that the connections $A$ and $\Theta$ must be $G_2$ instantons.

The vanishing of $\delta W$ with respect to the dilaton variations $\delta \phi$ gives
\[
(H+ h \varphi)\lrcorner\varphi - \frac{1}{2}\, \dd\varphi\lrcorner\psi = 0~.
\]
Then, using \eqref{eq:g2psi} we find
\be
\frac{1}{7}\, H\lrcorner\varphi + h - \frac{1}{2}\, \tau_0 = 0~.\label{eq:prepi1H}
\ee
Note that this implies that $W=0$ on the supersymmetric locus. 

The remaining terms in $\delta W$ correspond to the variations of the $G_2$ structure. Noting that
\[
*\delta\varphi = \frac{3}{7}\, \tr M\, \psi + i_m(\psi) - i_{\mathring M}(\psi)~,
\]
we write the remaining terms as
\begin{align*}
&\int_Y e^{-2\phi} \left\{
(- \dd\varphi + \dd\phi\wedge\varphi + h\, \psi)\wedge\delta\varphi
+ (H + h\, \varphi)\wedge\delta\psi\right\}
\\
&= \int_Y e^{-2\phi} \left\{
( - *\dd\varphi - \dd\phi\lrcorner\psi + h \, \varphi)\wedge *\delta\varphi
+ (H + h\, \varphi)\wedge\delta\psi\right\}
\\[10pt]
&= \int_Y e^{-2\phi} \left\{
\left[ \left(- \frac{3}{7}\, \tau_0 + h\right) \varphi 
+ \pi_{\bf 1}(H)\right]\wedge\psi\, \tr M
\right.
\\[5pt]
&\qquad\qquad\quad \left.
+ \left[ (3\tau_1 - \dd\phi)\lrcorner\psi + \pi_{\bf 7}(H) \right]\wedge i_m(\psi)
+ \left[\tau_3 + \pi_{27}(H)\right]\wedge i_{\mathring M}(\psi)
\right\}~.
\end{align*}
Hence
\[
\pi_{\bf 1}(H) =  \left(\frac{3}{7}\, \tau_0 - h\right) \varphi ~,
\quad
\pi_{\bf 7}(H)  = (- 3\tau_1 +  \dd\phi)\lrcorner\psi~,
 \quad
 \pi_{27}(H) = - \tau_3~. 
\]
The first of these equations together with \eqref{eq:prepi1H} give
\[
\pi_{\bf 1} (H) = \frac{1}{6} \, \tau_0~,\qquad h = \frac{1}{3}\, \tau_0
~,
\]
and the second, together with \eqref{eq:pretau1}
\[
\pi_{\bf 7}(H)  = -  \tau_1\lrcorner\psi~.
\]
Thus, both the internal and external flux is determined by the $G_2$ torsion.

Summarising our results, the critical points of $W$ are equivalent to requiring a heterotic $G_2$ system ({\it cf.} Section \ref{ssec:hetg2}) together with
\be
\tau_1 = \frac{1}{2}\, \dd\phi~, \qquad h = \frac{1}{3}\, \tau_0~. \label{eq:tors01}
\ee
That is, $\delta W = 0$ gives the supersymmetry conditions necessary for $\cN=1$ supersymmetry in three dimensions.  Moreover,  on the supersymmetric locus, $W = 0$.

Finally, let us remark briefly on the relation between the superpotential $W$ and the Hitchin functional 
\[
\int_Y \varphi \wedge\psi  \; ,
\]
which in the literature is sometimes referred to as a superpotential on the moduli space of $G_2$ holonomy manifolds \cite{2007arXiv0709.2987K}. 
The Hitchin functional \cite{Hitchin:2000jd} is a functional $\varphi$
whose critical points correspond to torsion-free $G_2$ structures  when restricted to closed $G_2$ structures on a fixed cohomology class. 
In this context it  has nothing to say about the gauge structure or the $G_2$ structures relevant to heterotic compactifications.   
The expectation, however, is that this functional, considered in the context of heterotic compactifications is related to the
metric on the moduli space of heterotic structures, as its form is very analogous to the corresponding $SU(3)$ structure functionals defined in the litterature \cite{Candelas:2016usb, Garcia-Fernandez:2018emx, Candelas:2018lib}. We leave this for future research.

\section{Conclusions and Outlook}
\label{sec:conc}

In this paper, we have proposed a superpotential $W$ for the three dimensional Yang--Mills supergravity that results from $\cN=1$ compactifications of the heterotic string on manifolds with $G_2$ structure in the large volume, weak string coupling limit.  These $\cN=1$ heterotic $G_2$ systems are three dimensional analogues of the four dimensional Hull--Strominger system.   There is a striking similarity between the three and four dimensional systems: both correspond to certain nilpotent operators on associated Atiyah-like bundles $\cal Q$, and the infinitesimal deformations are captured by certain $\cal Q$-valued cohomology classes. An important difference, however, is that the Hull--Strominger system only allows Minkowski vacua, whereas heterotic $G_2$ systems allow both Minkowski and AdS$_3$ vacua.  In this paper we have reported on some of the properties of these AdS$_3$ solutions, for example the relation between the curvature scales of the AdS$_3$ spacetime and $G_2$ structure manifold (see Appendix \ref{sec:apAdSKin}).  It would be interesting to use the leverage provided by the superpotential $W$ in more detailed studies of the low energy effective field theory resulting from heterotic $G_2$ systems, both around Minkowski and  AdS$_3$ vacua.

 Another important difference between heterotic $G_2$ systems and Strominger--Hull systems is the reduced amount of supersymmetry.  In particular, 3d, $\cN=1$ supergravity lacks nonrenormalisation theorems and so, in contrast to 4d $\cN=1$ supergravity, the superpotential may receive both perturbative and nonperturbative corrections. It follows that our results are valid only so long as we stay in the large volume, weak coupling limit. Even so, there are some signs that the string theory origins of $G_2$ systems may offer more protection than supersymmetry alone.  In particular, it has been argued \cite[Theorems 6 and 7]{delaOssa:2017pqy} that there is an equivalence, at all orders in $\alpha'$, between heterotic systems and the nilpotency of a certain differential operator, $\check{\cal D}$, whose cohomology is then related to the infinitesimal moduli space of heterotic $G_2$ systems. The crux of the matter is that, given an $\alpha'$ expansion, the heterotic Bianchi identity imposes relations between the order $n$ and $n+1$ terms, enabling a bootstrapping procedure in  $\alpha'$. This argument shows that corrections, perturbative in $\alpha'$, to the infinitesimal moduli space of heterotic systems are kept under control by the heterotic Bianchi identity. It should be pointed out, however, that the argument presumes the existence of an $\alpha'$ expansion, which is physically expected but not mathematically guaranteed, and in addition, is only valid for the infinitesimal moduli space. How this may change when we consider finite deformations is a very interesting direction for future research.

 The results of this paper may also be of interest for other heterotic configurations. For example,  we may restrict the torsion  of the heterotic $G_2$ structure and take a particular decompactification limit in order to connect to different types of supersymmetric \cite{Hull:1986kz,Strominger:1986uh} or supersymmetry-breaking four dimensional solutions of the heterotic string  \cite{Lukas:2010mf,Gray:2012md,deCarlos:2005kh,Gurrieri:2004dt,Gurrieri:2007jg,Klaput:2011mz,Klaput:2013nla,Klaput:2012vv}. A recent study of such connections between three and four dimensional hetorotic systems can be found in Ref.~\cite{delaOssa:2017gjq}. The superpotential $W$ provides a new tool that can be used to explore this relation, for example by studying how $W$ relates to the four dimensional superpotential \cite{Gurrieri:2004dt,delaOssa:2015maa,McOrist:2016cfl}.  Such a study is also expected to throw light on possible corrections of $W$. Another interesting avenue is to explore the relation with the three-dimensional Chern--Simons theories studies in Refs.~\cite{Gates:1991qn,Nishino:1991sr,Nishino:1991ej}.\footnote{However, note that the three-dimensional theory derived in section 7 of Ref.~\cite{Nishino:1991sr} is non-supersymmetric and arises from a compactification on a CY$_6 \times S^1$.}

The main motivation for our study is, however, the relevance of the superpotential for finite deformations of heterotic $G_2$ systems. Let us therefore end with a  discussion of higher order, and eventually finite, variations of the superpotential. We have shown, in Section \ref{sec:supotvar}, that  the first order variations of the superpotential $W$ vanish on a locus in field space that accurately reproduces the supersymmetry conditions. These supersymmetry constraints contain the conditions used to determine the infinitesimal moduli of heterotic $G_2$ systems \cite{delaOssa:2017pqy}. As a consequence, the second order variations of $W$ vanish on a locus in parameter space that corresponds to (a subspace of) the infinitesimal moduli of heterotic $G_2$ systems.

In more detail, it was shown in Ref.~\cite{delaOssa:2017pqy} that a generic deformation of the heterotic $G_2$ system can be thought of as a one-form with values in $\cal Q$, i.e. ${\cal X}\in\check\Omega^1({\cal Q})$, and that the deformation preserves the conditions of the heterotic $G_2$ system if and only if 
\begin{equation}
\label{eq:closedX}
\check{\cal D}{\cal X}=0\: ,
\end{equation}
where we refer to Section \ref{sec:G2diff} for the definitions of $\check{\cal D}$ and ${\cal Q}$.
Further, a generalised gauge transformation turns out to give rise to $\check{\cal D}$-exact one-forms, so the infinitesimal variations are represented by ${\cal X}\in H^1_{\check{\cal D}}({\cal Q})$ \cite{delaOssa:2017pqy}. In the same paper, it was argued that this structure is expected to persist to all orders in $\alpha'$, but it is less clear if the structure of higher order deformations will enjoy similar protection.

Although the results of Section \ref{sec:supotvar} guarantee that second order variations of the superpotential $W$ will reproduce these infinitesimal constraints on the variations, it is useful to sketch how this comes about. A variation of the superpotential  $W$, viewed as a functional of the fields $\Phi$, around the supersymmetric locus $\Phi = \Phi_0$, has the formal expansion 
\[
W[\Phi]=W[\Phi_0]+\Delta W = W[\Phi_0]+\delta W|_{\Phi_0} \delta\Phi + \tfrac12 \delta^2W |_{\Phi_0}\delta\Phi^2 +{\cal O}({\delta\Phi}^3) \; ,
\]
where the first two terms vanish by supersymmetry. Agreement with \eqref{eq:closedX} imposes that, to second order in the variations, we have up to a constant  
 \begin{equation}
\label{eq:DeltaW2}
\Delta W =\tfrac12\int_Me^{-2\phi_0}\,\langle{\cal X},{\cal D}{\cal X}\rangle\wedge\psi_0+{\cal O}({\cal X}^3)\:,
\end{equation}
where we rewrite the variation in terms of the $\cal Q$-valued one-form ${\cal X}$, and  $\langle\:,\:\rangle$ denotes the natural inner-product on $\cal Q$ (given by the metric $g_{\varphi_0}$ on $TY$ and the trace on $\cal G$). 
This expression is analogous to the finite variation of the superpotential for the Hull--Strominger system derived recently in Ref.~\cite{Ashmore:2018ybe}.

Some comments are in order concerning Eq.~\eqref{eq:DeltaW2}. One can wonder what the higher order deformations ${\cal O}({\cal X}^3)$ look like. In particular, we would like to understand whether there is some parametrisation of the finite deformations such that the higher order terms in the superpotential expansion truncate at finite order. In this regard, we recall the Hull--Strominger system where it was shown that a finite deformation of the superpotential could be parametrised as a Chern--Simons like cubic functional \cite{Ashmore:2018ybe}. This is also the natural expectation from a physical point of view, as cubic superpotential terms give rise to quartic terms in the four dimensional action. Such terms are exactly marginal in four dimensions, since there the mass dimension of a chiral field is 1.  The conditions that the superpotential and its variation vanish  implies that the deformation algebra can be described as an $L_3$-algebra. 

In three dimensional theories the chiral field has dimension 1/2, so it is the order six terms in the action that become marginal. These will derive from quartic terms in the superpotential. In analogy with the Hull--Strominger case, we might expect that a parametrisation of the deformations exist where the expansion of the superpotential in $\cal X$ truncates at quartic order. Similarly, it might also be the case that the corresponding deformation algebra now becomes an $L_4$ algebra. We hope to address these questions in the near future. 

Another interesting direction is to view $\Delta W$ as an action which can be quantised. Indeed, the famous Donaldson--Thomas invariants of Calabi--Yau manifolds equipped with holomorphic sheaves can be interpreted as correlation functions within holomorphic Chern--Simons theory \cite{donaldson1998gauge, thomas1997gauge, Thomas:1998uj}, and it has been suggested that similar invariants may be defined for the Hull--Strominger system using the corresponding superpotential \cite{Ashmore:2018ybe}. Donaldson and Segal have suggested that similar invariants could be defined for $G_2$ manifolds \cite{donaldson2009gauge}, though it has been debated to what extent such invariants depend on the given $G_2$ structure.\footnote{See e.g. \cite{Joyce:2016fij} and \cite{deBoer:2006bp} where the one-loop partition function of the $G_2$ analog of holomorphic Chern--Simons theory was computed.} The heterotic $G_2$ systems provide a string theoretic setting that includes configurations consisting of instanton bundles on manifolds with $G_2$ holonomy, which appear to be well suited for the study of invariants. Indeed, one might hope that the invariants defined using $\Delta W$ do not suffer from an anomalous dependence, as their definition would include an integration over the full parameter space of $G_2$ structures. We hope to test this conjecture in the future.

\section*{Acknowledgements}
The authors would like to express a special thanks to the Mainz Institute for Theoretical Physics (MITP) for its hospitality and support when this work was initiated.  We would also like to thank Anthony Ashmore, Mateo Galdeano Solans, and Marjorie Schillo for illuminating discussions. XD is supported in part by the EPSRC grant EP/J010790/1. The research of ML and MM is financed by the Swedish Research Council (VR) under grant number 2016-03873 and 2016-03503.  The work of ESS is supported by a grant from the Simons Foundation ($\#$488569, Bobby Acharya).

\begin{appendix}

 \section{Conventions} \label{ap:conv}

  We are interested in heterotic supergravity backgrounds of the form  $(M_3\times Y,\tilde{g}_3\oplus g_7)$ where $M_3$ is maximally symmetric and $(Y,g_7)$ is an integrable $G_2$-structure manifold (see Section \ref{ssec:hetg2} where we define a heterotic structure).  The physical three dimensional metric is a conformal rescaling of the restricted ten dimensional metric, $g_3:=e^{-n}\tilde{g}_3$ with the precise rescaling dictated by comparing the dimensional reduction with the canonical Einstein--Hilbert term (see Section \ref{sec:3dEHterm} for this calculation and discussion).

  In the next section, we specify the Clifford algebra conventions with an arbitrary conformal scaling. After that, in Appendix \ref{sec:apAdSKin}, we give our definition of the kinetic operator for the three dimensional gravitinos, which is subtle when we have non-zero cosmological constant.  
  
    Throughout this paper, three dimensional indices will be denoted $\mu,\nu,\ldots$, while seven dimensional indices are indicated by $i,j,\ldots$; ten dimensional coordinates are labeled by $(X^{M})\rightarrow (x^\mu,y^i)$.  A bar on the indices indicates that they are flat space indices.

\subsection{Clifford algebra}\label{sec:appCliff}
     Starting with the flat space 10d Clifford algebra, $\Gamma^{\bar{M}}$, under the group decomposition $SO(1,9)\hookleftarrow SO(1,2)\times SO(7)$ we have the Clifford algebra decomposition\footnote{The subscripts in parentheses are to explicitly keep track of which algebra the $\Gamma$ matrices are in. They will not appear in the main text.} \cite{Lukas:2010mf}:
  \begin{align}
    \Gamma^{\bar{\mu}}_{(10)}&=\Gamma^{\bar{\mu}}_{(3)}\otimes\Id\otimes\sigma^{2}\\
    \Gamma_{(10)}^{\bar{i}}&=\Id\otimes\Gamma_{(7)}^{\bar{i}}\otimes\sigma^1\,,
  \end{align}
 where $\sigma^1$ and $\sigma^2$ are the Pauli matrices
  \begin{align}
    \sigma^1&=\left(\begin{array}{cc}0&1\\1&0\end{array}\right)\\
    \sigma^2&=\left(\begin{array}{cc}0&-i\\i&0\end{array}\right)\,.
  \end{align}
  We will assume that the ten dimensional matrices are pure imaginary and as a consequence we choose the $\Gamma_{(7)}$ to be purely imaginary matrices, while the $\Gamma_{(3)}$ are real.  This choice is made to be consistent with the conventions of \cite{Lukas:2010mf}.

This algebra decomposition induces a Clifford bundle decomposition on the space $M_3\times  Y$. Indeed, let $E^{\bar{M}}_M$ be the ten dimensional vielbein, $E^{\bar{M}}_M\,E^{\bar{N}}_N\,\eta_{\bar{M}\bar{N}}=g^{(10)}_{MN}$. It has the decomposition
 \begin{align}
 E^{\bar{M}}_M&\longrightarrow(e^{n/2}\,\tilde{e}^{\bar{\mu}}_\mu,E^{\bar{i}}_i)\,,\label{eq:vielbein}
 \end{align}
 where $e^{\bar{\mu}}_\mu$ is the vielbein associated to $g_3$.
 
  This decomposition gives the relationship between curved space Clifford matrices:
  \begin{align}
    \Gamma^\mu_{(10)}=E^\mu_{\bar{\mu}}\,\Gamma^{\bar{\mu}}_{(10)}&=e^{-n/2}(e^\mu_{\bar{\mu}}\,\Gamma^{\bar{\mu}}_{(3)}\otimes\Id\otimes\sigma_2)\notag\\
                                                            &=e^{-n/2}(\Gamma^\mu_{(3)}\otimes\Id\otimes\sigma_2)\label{eq:expg3}\\
    \Gamma^i_{(10)}=E^i_{\bar{i}}\,\Gamma^{\bar{i}}_{(10)}&=(\Id\otimes\Gamma^i_{(7)}\otimes\sigma_1)\,.\label{eq:expg7}
  \end{align}
To keep the scaling factors associated with Clifford matrices distinct from other metric contributions, we introduce 
\be
\alpha:=-n/2
\ee
as the factor in the relation between ten and three dimensional Clifford matrices.

  Next, we explicitly realise the isomorphism $\mathcal{S}_{10}\cong\mathcal{S}_3\otimes\mathcal{S}_7$, where $\mathcal{S}_{10}$ is a ten dimensional Majorana--Weyl spinor space.  Given a three dimensional spinor, $\rho$, and seven dimensional spinor, $\lambda$, we write the ten dimensional spinor as $\rho\otimes\lambda\otimes\theta$, where $\theta$ is a definite chirality two dimensional spinor specifying the ten dimensional chirality.  We will only be interested in ten dimensional positive-chirality gravitinos, $\Psi_M$ with decomposition:
  \begin{equation} \Psi_{\bar{M}}\sim \left\{\begin{array}{cc}\rho_{\bar{\mu}}\otimes\lambda\otimes\theta& \mu=M=0,1,2\\\rho\otimes\lambda_{\bar{i}}\otimes\theta&\quad i=M=3,\ldots,10\end{array}\right.\,,
  \end{equation}
  with $\theta$ a positive-chirality two dimensional spinor. In this paper, only the three dimensional gravitinos are relevant and we will have no cause to consider the seven dimensional gravitino. Furthermore, in order for the three dimensional gravitino to be light, relative to the compactification scale, we require that the seven dimensional fermion, $\lambda$, be the spinor defining the $G_2$ structure (see Section \ref{ssec:g2struct} for more details).
  
  However, we have to be a little bit careful in correctly identifying the three dimensional gravitino since the conformal factors we have introduced induce scaling here, too.  We can deduce the correct factor by observing that $\Gamma_\mu\psi$ and $\rho_\mu$ are in the same ten dimensional supermultiplet, where $\psi$ is the dilatino. As a consequence, we find that the correct decomposition to the three dimensional gravitino is:
 \begin{align}
    \Psi_\mu&\longrightarrow  e^{n/2}(\rho_\mu\otimes\lambda\otimes\theta)\,.
  \end{align}
   
 As we did for the Clifford algebra above, we introduce an extra variable
  \be
  \beta:=n/2
  \ee
  in order to keep the origin of each factor explicit.\footnote{It can be checked that with this definition, the three dimensional gravitino kinetic term obtained from dimensional reduction is canonically normalised, so long as the Einstein--Hilbert term is.}

  Gathering the above, we have the following conventions:
  \begin{align}
    g_{10}&=e^n\,g_{3}\oplus g_7\label{eq:metcons}\\
    \Gamma_{(10)}^\mu&=e^\alpha\,\Gamma_{(3)}^\mu\otimes\Id\otimes\sigma_2\label{eq:3clifcons}\\
    \Gamma_{(10)}^i&=\Id\otimes \Gamma_{(7)}^i\otimes\sigma_1\label{eq:7clifcons}\\
    \Psi_\mu&=e^\beta(\rho_\mu\otimes\lambda\otimes\theta)\label{eq:fermcons}\\
    \bar{\Psi}_\mu&=e^{\beta+\alpha}(\bar{\rho}_\mu\otimes\lambda^\dagger\otimes\theta^\dagger\sigma_2)\,,\label{eq:fermcons2}
  \end{align}
  where we should recall that ${\alpha}={-n/2}=-\beta$. We note that in the final equation, the $\Gamma^0$ in $\bar{\Psi}=\Psi^\dagger\Gamma^0$ adds the $\alpha$ contribution; although $\alpha+\beta= 0$, keeping every conformal scale factor explicit makes the dimensional reduction more transparent.

  \subsection{AdS$_3$ kinetic terms and curvature scales}\label{sec:apAdSKin}

 We are studying backgrounds in which the maximally symmetric space $M_3$ is not necessarily Minkowski.  Since a non-trivial cosmological constant gives a mass-like source term in the Killing spinor equations, we must be careful about identifying the physical mass terms as opposed to mass-like kinetic contributions. We will deal with this subtlety by absorbing the cosmological constant into a redefined covariant derivative, $\nabla_\mu$, that annihilates the Killing spinors. This operator can be identified as the correct physical kinetic operator, in the sense that its eigenvalues correspond to the  physical mass. In particular, with this convention supersymmetric solutions have a massless gravitino and vanishing superpotential.  The point of view pursued here was learnt from Cecotti's book, \cite[Ch.6]{cecotti_2015}.  The first thing we will do is identify the precise cosmological constant and Killing spinor equation, then use this to identify the correct kinetic operator. Secondly, we will discuss the curvature scales of the heterotic AdS$_3$ solutions.
  
  In AdS$_3$ space, the Killing spinor equation is (see e.g. \cite{2009PThPh.122..631K}):
  \begin{align*}
    \DD_\mu^{(3)}\rho&=\frac{a}{2}\Gamma_\mu\rho\,,
  \end{align*}
  where $a$ is related to the cosmological constant, $\Lambda$, via $\Lambda=-2a^2$, and can be determined from the ten dimensional Killing spinor equation:
  \begin{align}
    (\DD_M-\frac{1}{8}H_{MNP}\Gamma^{NP})\epsilon&=0\,.\label{eq:10dKSE}
  \end{align}
Using the Clifford algebra decomposition given in Appendix \ref{sec:appCliff} we  write $\epsilon=\rho\otimes\eta\otimes\theta$ and compute:
  \begin{align*}
    &(\DD_\mu-\frac{1}{8}H_{\mu\nu\kappa}\Gamma^{\nu\kappa})\epsilon=0\\
    \implies&(\DD_\mu-\frac{1}{8}H_{\mu\nu\kappa}(e^{2\alpha}\,\Gamma_3^{\nu\kappa}\otimes\Id\otimes\Id))(\rho\otimes \eta\otimes\theta)=0\\
   \implies&(\DD_\mu-\frac{1}{8}e^{2\alpha}\,\Gamma^{\nu\kappa}_3H_{\mu\nu\kappa})\rho=0\,.
  \end{align*}

The two $\Gamma$-matrices are related to a single $\Gamma$-matrix via the three dimensional $\Gamma$-duality that can be found in Polchinski \cite[App. B]{Polchinski:1998rr}:
  \begin{align*}
    \Gamma^{\nu\kappa}&=\epsilon^{\nu\kappa\sigma}\Gamma_\sigma\sqrt{-g_3}\,.
  \end{align*}
  It follows that:
  \[
    \Gamma^{\nu\kappa}H_{\mu\nu\kappa}=\sqrt{-g_3}\,\epsilon^{\nu\kappa\sigma}\,H_{\nu\kappa\mu}\,\Gamma_\sigma
    =2*_3H_{(3)}\,\Gamma_\mu
 \]
 where we have recognised 
 $
 \epsilon^{\nu\kappa\sigma}H_{\nu\kappa\mu}=2\delta^\sigma_\mu *_3H_{(3)}$. 
 Hence the Killing spinor equation becomes:
  \begin{align}
    \DD_\mu\rho&=\frac{1}{4}*_3H_{(3)}\,e^{2\alpha}\,\Gamma_\mu\rho\,.
  \end{align}
  Therefore, we must have 
  \[a=\frac{1}{2}*_3H_{(3)} e^{2\alpha}\,.\]

  With $a$ now identified we can define the physical three dimensional covariant derivative
  \begin{align}
\nabla_\mu:=\DD_\mu-  \frac{1}{4}\, *_3H_{(3)}\,  e^{2\alpha}\, \Gamma_\mu\,.\label{eq:apAdSCovDer}
  \end{align}

  Note that this shift really does contribute a mass-like term, since the kinetic operator $\Gamma^{\mu\nu\kappa}\,\nabla_\nu\Psi_\kappa$ includes the extra term proportional to:
  \[
    \Gamma^{\mu\nu\kappa}\Gamma_\nu
                                   =\epsilon^{\mu\nu\kappa}\Gamma_\nu
    =-\Gamma^{\mu\kappa}\,,
\]
   where we have used the three dimensional $\Gamma$-matrix duality twice.  This can also be computed using the GAMMA package \cite{Gran:2001yh}.

  As a consequence, the kinetic term in the 3d action is:
  \be 
   \bar{\rho}_\mu\Gamma^{\mu\nu\kappa}\nabla_\nu\rho_\kappa
   =\bar{\rho}_\mu\Gamma^{\mu\nu\kappa}\DD_\nu\rho_\kappa+
   \frac{1}{4}\, *_3H_{(3)}\,e^{2\alpha}\,\rho_\mu\Gamma^{\mu\kappa}\rho_\kappa\,.\label{eq:3dkinAdS}
  \ee
  Comparing to equation (\ref{eq:RSeqn}) gives the na\"ive mass interpretation of the extra term in (\ref{eq:apAdSCovDer}). 

  Finally, we note that although $*_3H_{(3)}$ is a natural object from the three dimensional perspective, our results are simplified by introducing the scalar field, $f=* H_{(3)}$ where the Hodge star is taken with respect to the restriction of the ten dimensional  metric, i.e. without the conformal factor.  This is related to $*_3H_{(3)}$ as follows:
  \be 
  \label{eq:fstarH}
  *_3H_{(3)} =\frac{\sqrt{-g_3}}{3!}H^{\mu\nu\lambda}\epsilon_{\mu\nu\lambda}
              =\frac{\sqrt{-g_3}}{3!}f\varepsilon_{\rho\sigma\kappa}g^{\mu\rho}_3g^{\nu\sigma}_3g^{\kappa\lambda}_3\epsilon_{\mu\nu\lambda}
            = \frac{\sqrt{-g_{10}|_3}}{\sqrt{-g_3}}\, f
  =e^{3n/2}f\,.
  \ee
  In particular, since $2\alpha=-n$, we find that $a=\frac{1}{2}f e^{n/2}$ and the three dimensional cosmological constant is
  \be \label{eq:3dcc}
    \Lambda=-\frac{1}{2}\,f^2\,e^n\,.
  \ee

  Using the fact that $f$ is proportional to the internal torsion component $\tau_0$, \eqref{eq:tors01}, we can observe that the AdS$_3$ curvature scale is set by $\tau_0^2$.  On the other hand, the internal curvature depends on all components of the torsion, \cite{Bryant:2005mz}. The scalar curvature, for instance, is given by\footnote{Here, we use notation $|\tau|^2:=\tau\lrcorner\tau$}:
  \begin{align*}
    \cal R &=12d^\dagger\tau_1+\frac{21}{8}\tau_0^2+30|\tau_1|^2-\frac{1}{2}|\tau_2|^2-\frac{1}{2}|\tau_3|^2\,;
  \end{align*}
  in particular, the scalar curvature for an $\cN=1$ heterotic system is:
  \begin{align}
    \cal R&=6\Delta_7\phi-\left(\tfrac{21}{2}\right)^3e^{-n}\Lambda+\tfrac{30}{4}|d_7\phi|^2-\tfrac{1}{2}|\pi_{\mathbf{27}}H|^2\,,
  \end{align}
  where $\Delta_7$ is the seven dimensional Laplacian defined with the $G_2$ metric. 
  There exist comparable expressions for the Riemann tensor and Ricci tensor.

  Since the different components of the torsion decouple, at least infinitesimally, one could hope to tune the internal torsion and produce a scale separated AdS$_3$ compactification.  We leave this possibility to explore in future.

\end{appendix}

\newpage
\bibliographystyle{JHEP}

\bibliography{bibliography}

\end{document}